\documentclass[prl,twocolumn,longbibliography,superscriptaddress,preprintnumbers]{revtex4-2}
\usepackage{comment}
\usepackage{bm}
\usepackage{dsfont}
\usepackage{graphicx}
\usepackage{physics}
\usepackage{amsfonts}
\usepackage{epstopdf}
\usepackage{balance}
\usepackage[dvipsnames]{xcolor}
\usepackage{calc}
\usepackage{natbib}
\usepackage[colorlinks,
            linkcolor=blue,
            anchorcolor=blue,
            citecolor=blue,
            urlcolor=blue]{hyperref}
\usepackage{lipsum}
\usepackage[version=3]{mhchem} 
\usepackage{color}
\usepackage{soul}
\usepackage[export]{adjustbox}
\usepackage{makecell}
\usepackage{multirow}
\usepackage{tabularx,array,diagbox}
\usepackage{booktabs}

\makeatletter
\renewcommand{\maketag@@@}[1]
{\hbox{\m@th\normalsize\normalfont#1}}%
\makeatother
  
\begin{document}

\title{\resizebox{\textwidth}{!}{Supermoir\'e Reconstruction and Topological Mosaics in Twisted Trilayer WSe$_2$ and MoTe$_2$}}

\author{Hai Meng}
%\email{menghai@whu.edu.cn}
\affiliation{School of Physics and Technology, Wuhan University, Wuhan 430072, China}
\author{Yang Xu}
%\email{yang.xu@iphy.ac.cn}
\affiliation{Beijing National Laboratory for Condensed Matter Physics, Institute of Physics, Chinese Academy of Sciences, Beijing 100190, China}
\affiliation{School of Physical Sciences, University of Chinese Academy of Sciences, Beijing 100049, China}
\author{Fengcheng Wu}
\email{wufcheng@whu.edu.cn}
\affiliation{School of Physics and Technology, Wuhan University, Wuhan 430072, China}
\affiliation{Wuhan Institute of Quantum Technology, Wuhan 430206, China}

\begin{abstract}
We investigate lattice relaxation and band structures of helical and alternating twisted trilayer WSe$_2$ and MoTe$_2$ using machine-learning force fields and large-scale \emph{ab initio} calculations. Interference between the two bilayer moir\'e lattices generates a supermoir\'e lattice that, upon relaxation, reconstructs into a few dominant domain types with locally commensurate bilayer moir\'e lattices. Because the systems lack $C_{2z}$ symmetry, domains otherwise related by this symmetry become energetically and topologically distinct, unlike in twisted trilayer graphene. Band structure calculations show that the topmost valence bands originate from the $K$ ($K'$) valleys and carry domain-dependent valley Chern numbers. The resulting supermoir\'e lattice hosts a mosaic of topologically inequivalent domains, offering a platform for exploring correlated and topological physics.
\end{abstract}
\maketitle

\textit{Introduction.—}
Moir\'e materials have become an important platform for studying tunable electronic phases. Beyond graphene-based systems~\cite{lopes2007graphene,rafi2011moire,cao2018correlated,cao2018unconventional}, transition metal dichalcogenides (TMDs), with finite effective mass, strong spin-orbit coupling, and spin-valley locking~\cite{xiao2012coupled,xu2014spin}, provide another prototypical moir\'e material~\cite{wu2017topological,wu2018hubbard,naik2018ultraflatbands,wu2019topological,pan2020band,devakul2021magic}, hosting correlated insulating states~\cite{regan2020mott,tang2020hubbard,xu2020correlated,wang2020correlated}, integer and fractional quantum anomalous Hall insulators~\cite{li2021qahe,cai2023signatures,zeng2023thermodynamic,park2023observation,xu2023observation,foutty2024mapping}, and unconventional superconductivity~\cite{xia2025superconductivity,guo2025superconductivity,xia2026bandwidth,guo2026angle}.
Twisted trilayer TMDs offer an even richer platform, with theoretical studies predicting stacking-tunable electronic structures~\cite{albuhairan2023band,choi2025higher,fedorko2025engineer,nakatsuji2025moire,liang2025moire}, though these rely on approximate continuum models with phenomenologically fitted parameters. The twist angle pair $(\theta_{12},\theta_{23})$, between layers $1$--$2$ and layers $2$--$3$, gives rise to helical ($\theta_{12}\theta_{23}>0$) and alternating ($\theta_{12}\theta_{23}<0$) configurations, both extensively studied in twisted trilayer graphene~\cite{zhu2020twisted,zhang2021correlated,park2021tunable,hao2021electric,devakul2023magic,kim2023imaging,popv2023magic,nakatsuji2023multiscale,xia2025topological,hoke2026imaging}, motivating analogous studies here. For general twist angles, the two bilayer moir\'e lattices formed by adjacent layers differ in periodicity and/or orientation, and their interference gives rise to a supermoir\'e (moir\'e-of-moir\'e) structure, where lattice relaxation occurs on both the moir\'e and supermoir\'e length scales~\cite{devakul2023magic,nakatsuji2023multiscale,guerci2024chern}. As demonstrated in studies of twisted bilayer TMDs~\cite{wang2024fractional,wang2025higher,xu2025multiple}, a full \emph{ab initio} treatment is essential to capture this lattice reconstruction and its impact on electronic structure.

In this Letter, we perform large-scale \emph{ab initio} calculations for both helical and alternating twisted trilayer moir\'e WSe$_2$ and MoTe$_2$. We find that supermoir\'e trilayers with closely matched $|\theta_{12}|$ and $|\theta_{23}|$ relax into large domains, within each of which the two bilayer moir\'e lattices become commensurate, with a domain-dependent in-plane shift. The topmost valence moir\'e bands in these domains, derived from the $K$ ($K'$) valleys, can possess nontrivial band topology, with different domains carrying distinct valley Chern numbers. Unlike twisted trilayer graphene, where $C_{2z}$ (twofold rotation around the out-of-plane $z$ axis) symmetry renders symmetry-related domains equivalent, twisted trilayer TMDs lack this symmetry, so the dominant domains can become inequivalent, a feature not captured by available continuum approximations~\cite{nakatsuji2025moire}. The resulting supermoir\'e structures can be viewed as mosaic tiles of domains with varied valley (locked to spin) Chern numbers, providing a microscopic foundation for investigating correlated and topological physics in twisted trilayer TMDs.

\textit{Atomistic modeling.—}
To describe lattice relaxation in twisted trilayer TMDs, we perform molecular dynamics (MD) simulations using machine-learning force fields (MLFFs) trained on \emph{ab initio} data. At small twist angles, local atomic stacking in a trilayer can be approximated by a non-twisted trilayer with a relative in-plane shift, so we use non-twisted trilayer structures as training data. Configurations are parametrized by $\{\bm{\delta}_b,\bm{\delta}_t\}$, where $\bm{\delta}_{s}=\mu_s\bm{a}_1+\nu_s\bm{a}_2$ ($s=b,t$) is the shift of layer 1 (bottom) or layer 3 (top) relative to layer 2 (middle), with $\bm{a}_1=a(1,0)$ and $\bm{a}_2=a(1/2,\sqrt{3}/2)$ being the monolayer lattice vectors. To span the atomic environments occurring in a twisted trilayer supermoir\'e lattice, we sample $\mu_s,\nu_s\in\{-1/2+k/6\mid k=0,\dots,6\}$ (49 shifts per interface); accounting for the equivalence of $\{\bm{\delta},\bm{\delta}'\}$ and $\{\bm{\delta}',\bm{\delta}\}$ by mirror reflection through layer 2, this gives 1225 unique training configurations. For each, we build a $3\times3$ supercell, apply random atomic displacements ($|p_\alpha|\leq0.15$ \text{\AA}) and a random lattice rescaling $\lambda\in[0.95,1.05]$.

MLFFs are trained on this 1225-configuration \emph{ab initio} dataset by minimizing a loss function combining energy, force, and virial terms, using DeePMD-kit~\cite{wang2018dpmd1,zeng2023dpmd2}; \emph{ab initio} data are computed with ABACUS~\cite{chen2010abacus1,li2016abacus2}. The relaxed structures are obtained via MD simulations in LAMMPS~\cite{plimpton1995lammps1,thompson2022lammps2}, using the obtained MLFFs (see Supplemental Material (SM)~\cite{SM} for more details). 

\begin{figure}[t]
    \centering
    \includegraphics[width=1.\columnwidth]{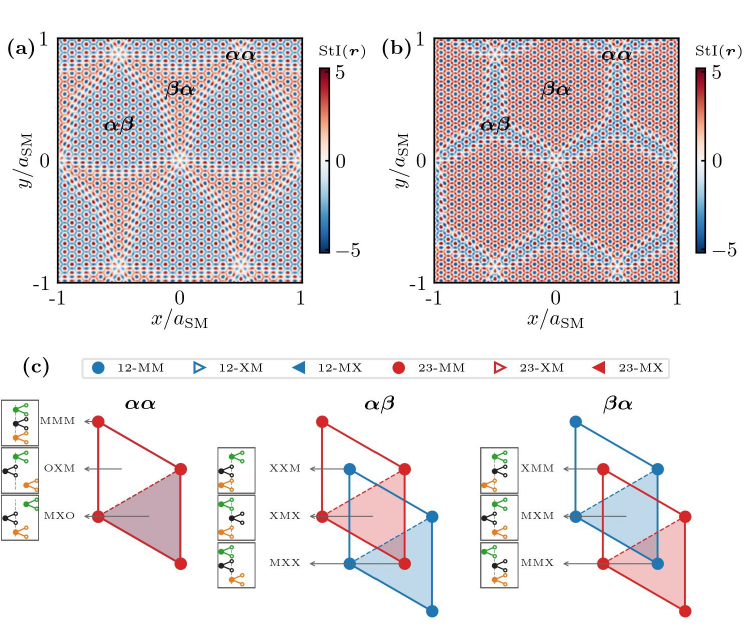}
    \caption{(a,b) Spatial pattern of relaxed htWSe$_2$ and htMoTe$_2$ with $\theta_{12}=\theta_{23}=3.89^\circ$, respectively. $a_{\text{SM}}$ denotes the supermoir\'e period. (c) Schematic illustration of $\alpha\alpha$, $\alpha\beta$, and $\beta\alpha$ domains. Blue and red rhombi indicate the moir\'e lattice of layers $1$--$2$ and $2$--$3$, respectively. Insets show local high-symmetry stackings, with orange, black, and green circles denoting atoms in layers 1, 2, and 3, respectively. Solid and empty circles represent metal (M) and chalcogen (X) atoms.} 
    \label{fig:hel_stru}
\end{figure}

\textit{Supermoir\'e relaxation.—}
When $|\theta_{12}|$ and $|\theta_{23}|$ are close, the moir\'e periods of the $1$--$2$ and $2$--$3$ layer pairs are close, giving rise to a supermoir\'e lattice. We investigate the supermoir\'e relaxation in two representative structures, one helical and one alternating, with twist angle pairs $(\theta_{12},\theta_{23})=(3.89^\circ,3.89^\circ)$ and $(2.20^\circ,-1.84^\circ)$, respectively. In the first case, $3.89^\circ$ corresponds to a commensurate twist angle in twisted bilayers. Because $\theta_{12}=\theta_{23}=3.89^\circ$, the $1$--$2$ and $2$--$3$ moir\'e superlattices share the same periodicity and are themselves twisted by a commensurate angle of $3.89^\circ$. Consequently, the resulting supermoir\'e structure is commensurate.
The second case, $(\theta_{12},\theta_{23})=(2.20^\circ,-1.84^\circ)$, represents a nearly commensurate configuration. The two moir\'e superlattices have different periods and a small relative twist of $0.18^\circ$, which would generate an incommensurate supermoir\'e pattern with a period too large for practical computation. By slightly modifying the lattice constants of layers 1 and 2 by factors of $0.9998$ and $1.0005$, respectively, the residual relative twist between the two moir\'e lattices is eliminated, yielding a commensurate structure with a $5/6$ ratio between the periods of the $1$--$2$ and $2$--$3$ moir\'e lattices \cite{SM}. This tiny lattice adjustment preserves the essential local stacking configurations while enabling efficient numerical simulations. 

\begin{figure}[t]
    \centering
    \includegraphics[width=1.\columnwidth]{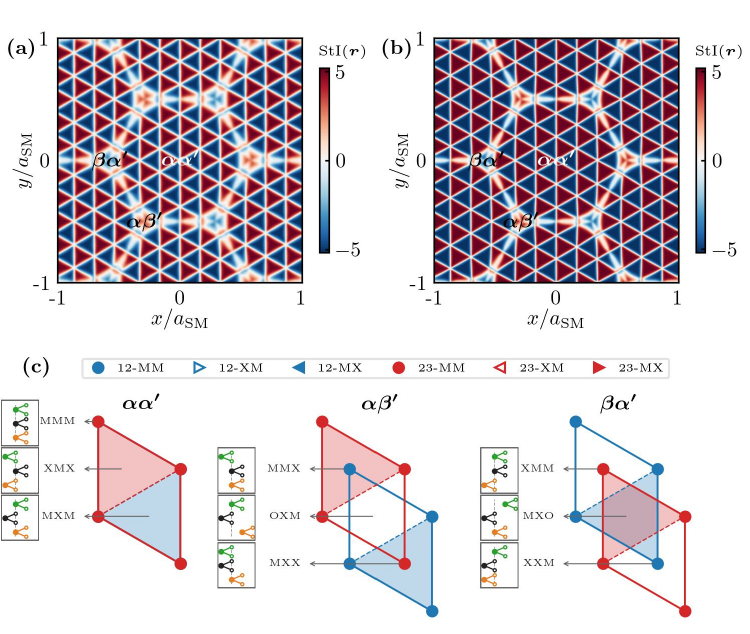}
    \caption{(a,b) Spatial pattern of relaxed atWSe$_2$ and atMoTe$_2$ with $(\theta_{12},\theta_{23})=(2.20^\circ,-1.84^\circ)$, respectively. (c) Schematic illustration of $\alpha\alpha'$, $\alpha\beta'$, and $\beta\alpha'$ domains.} 
    \label{fig:alt_stru}
\end{figure}

We relax both supermoir\'e structures using the atomistic modeling procedure, and characterize the resulting local stacking configurations through the stacking identity (StI), defined as
\begin{align}
\label{eq:sti}
&\text{StI}(\bm{r})=-\sum_{\epsilon=1,3}\sum_{j=1}^3\sin{[\bm{b}_j\cdot\bm{\delta}_{\epsilon}(\bm{r})]},\notag\\
&\bm{b}_j=\frac{4\pi}{\sqrt{3}a}(\cos{\frac{4j-5}{6}\pi},\sin{\frac{4j-5}{6}\pi}),
\end{align}
where $\bm{\delta}_{\epsilon}(\bm{r})$ is the local in-plane shift of layer $\epsilon$ relative to layer $2$ at position $\bm r$, and  $\bm{b}_j$ are the reciprocal lattice vectors of layer $2$. Here we use $\text{StI}(\bm{r})$, which encodes interlayer shifts $\bm{\delta}_1(\bm{r})$ and $\bm{\delta}_3(\bm{r})$ of both interfaces into sinusoidal functions, rather than the commonly used interlayer distances, to better resolve the local stackings.  

We first investigate the relaxed supermoir\'e structures of helical trilayer WSe$_2$ (htWSe$_2$) and MoTe$_2$ (htMoTe$_2$), shown in Figs.~\ref{fig:hel_stru}(a) and \ref{fig:hel_stru}(b), respectively. The function $\text{StI}(\bm{r})$ is displayed using a diverging red-blue color scale, where red (blue) denotes positive (negative) StI, with the darkest values corresponding to XMX (MXM) stacking, which are illustrated in Fig.~\ref{fig:hel_stru}(c).

There are three characteristic domain types in the supermoir\'e structures. The $\alpha\beta$ domain consists of dark-red XMX regions forming a triangular lattice in a blue background, while the $\beta\alpha$ domain shows the opposite pattern, with dark-blue MXM regions embedded in a red background. The $\alpha\alpha$ domain appears as a hexafoil-shaped motif with alternating light-red and light-blue petals. In all three domains, the $1$--$2$ and $2$--$3$ moir\'e lattices tend to stack commensurately with a relative in-plane shift, as depicted in Fig.~\ref{fig:hel_stru}(c).

\begin{figure}[t]
    \centering
    \includegraphics[width=1.\columnwidth]{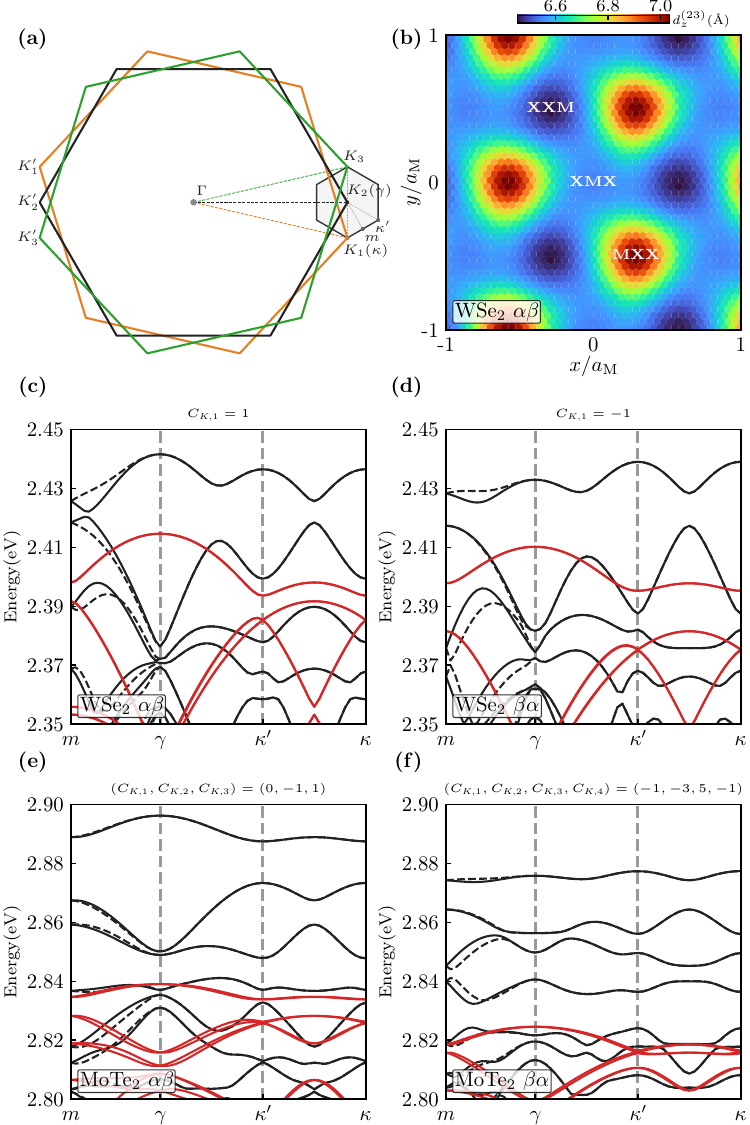}
    \caption{(a) Moir\'e Brillouin zone (gray hexagon) of relaxed helical trilayer TMDs in commensurate domains. Orange, black, and green hexagons denote the Brillouin zones of layers 1, 2, and 3, respectively. (b) Metal-metal vertical distance $d_z^{(23)}$ between layers 2 and 3 in the relaxed $\alpha\beta$ domain of htWSe$_2$.  $a_{\text{M}}$ denotes the moir\'e period. (c-f) Moir\'e valence bands at the $K$ (black solid), $K^\prime$ (black dashed), and $\Gamma$ (red) valleys of helical trilayer TMDs. $C_{K,i}$ denotes the Chern number of the $i$-th valence band at the $K$ valley. Panels (c–f) correspond to htWSe$_2$ in the $\alpha\beta$ (c) and $\beta\alpha$ (d) domains, and htMoTe$_2$ in the $\alpha\beta$ (e) and $\beta\alpha$ (f) domains. In (b)-(f), the lattice constant of layer $2$ is slightly expanded and $\theta_{12}=\theta_{23}=3.67^\circ$.} 
    \label{fig:hel_band}
\end{figure}

\begin{figure}[t]
    \centering
    \includegraphics[width=1.\columnwidth]{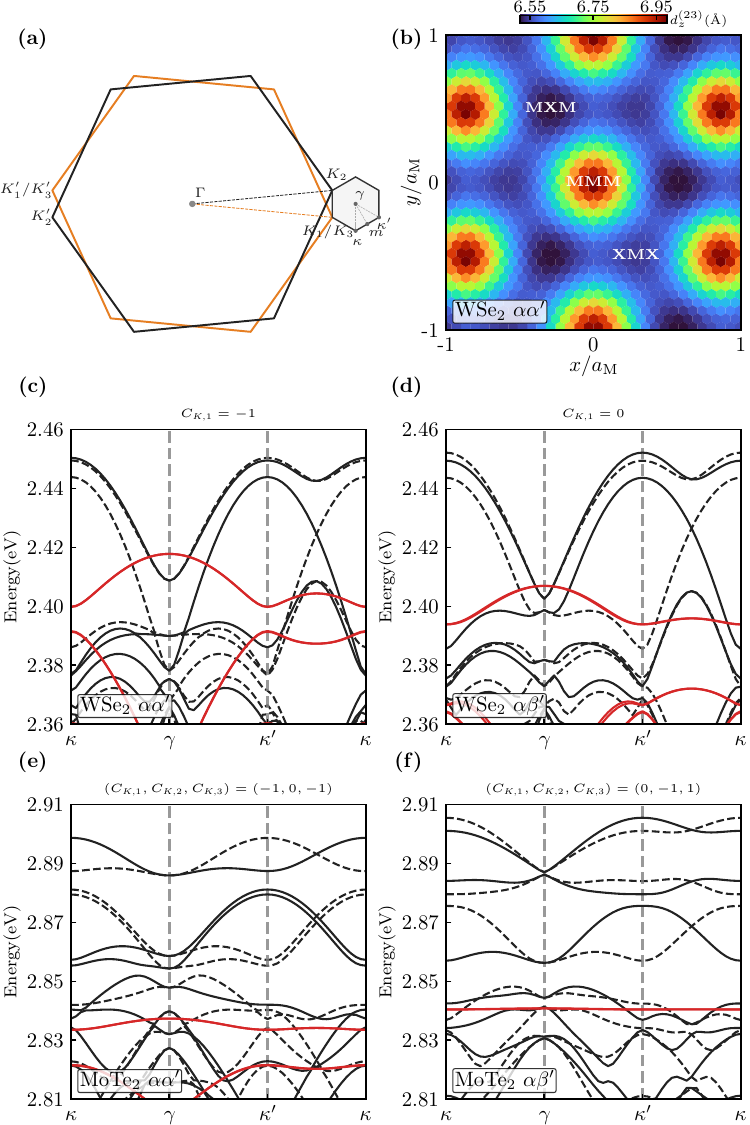}
    \caption{(a) Moir\'e Brillouin zone (gray hexagon) of alternating trilayer TMDs with  $\theta_{12}=-\theta_{23}$. Orange and black hexagons denote the Brillouin zones of layer 1 (3) and layer 2, respectively. (b) Interlayer distance $d_z^{(23)}$ between layers 2 and 3 in the relaxed $\alpha\alpha^\prime$ domain of atWSe$_2$.  $a_{\text{M}}$ denotes the moir\'e period. (c-f) Moir\'e valence bands at the $K$ (black solid), $K^\prime$ (black dashed), and $\Gamma$ (red) valleys of alternating trilayer TMDs.  Panels (c–f) correspond to atWSe$_2$ in the $\alpha\alpha^\prime$ (c) and $\alpha\beta^\prime$ (d) domains, and atMoTe$_2$ in the $\alpha\alpha^\prime$ (e) and $\alpha\beta^\prime$ (f) domains. In (b)-(f), $\theta_{12}=-\theta_{23}=3.89^\circ$.} 
    \label{fig:alt_band}
\end{figure}

The competition among these local domains drives a substantial reconstruction of the supermoir\'e lattice. The $\alpha\alpha$ domain shrinks considerably, while the domain walls between $\alpha\beta$ and $\beta\alpha$ become bent, reflecting their energetic imbalance. This asymmetry arises because these two domains are not symmetry-related in helical trilayer TMDs due to the absence of $C_{2z}$ symmetry, in contrast to helical trilayer graphene where they are related by $C_{2z}$ symmetry~\cite{devakul2023magic,nakatsuji2023multiscale}.

In htWSe$_2$, the domain walls bend from $\alpha\beta$ toward $\beta\alpha$, indicating that $\alpha\beta$ domains are energetically favored; in htMoTe$_2$, the bending runs the opposite way, favoring $\beta\alpha$ domains instead. This reversal reveals a subtle energy difference between the two domains \cite{SM}.
The bending is stronger in htMoTe$_2$ than in htWSe$_2$, which can be attributed to the mechanically softer lattice of MoTe$_2$~\cite{zeng2015electronic}.

For the alternating trilayer WSe$_2$ (atWSe$_2$) and MoTe$_2$ (atMoTe$_2$), their relaxed supermoir\'e structures are displayed in Figs.~\ref{fig:alt_stru}(a) and \ref{fig:alt_stru}(b), respectively. The patterns are qualitatively similar for both materials but markedly different from the helical case. The dominant $\alpha\alpha^\prime$ domain forms large expanded hexagons, within which red (positive StI) and blue (negative StI) triangles indicate XMX and MXM stackings, respectively. The $\alpha\beta^\prime$ and $\beta\alpha^\prime$ domains, in contrast, are reduced to trilobed features at the hexagon corners. This expansion of $\alpha\alpha^\prime$ and shrinkage of the other domains imply that $\alpha\alpha^\prime$ is energetically preferred. Comparing the two materials, atMoTe$_2$ exhibits stronger moir\'e relaxation than atWSe$_2$, as seen from the thinner domain walls in Fig.~\ref{fig:alt_stru}(b).
These results, together with those for the helical trilayers, suggest a general trend: domains containing XMX and/or MXM tend to be energetically favored.

\textit{Moir\'e band structure.—}
We examine the electronic band structures of the dominant domains generated by lattice relaxation, each of which locally resembles a moir\'e lattice with a well-defined period.
For each domain type, we choose a commensurate structure, relax it via MD using the MLFF, and finally obtain the electronic Hamiltonians of the relaxed structures from \emph{ab initio} calculations. We adopt the truncated atomic plane wave (TAPW) method~\cite{miao2023tapw} to calculate the moir\'e band structures and band topology, focusing on the low-energy states near the valence band maximum, which are derived from the $K$, $K^\prime$, and $\Gamma$ valleys and thus allow projection of the Hamiltonians onto these valleys (see SM~\cite{SM} for calculation details and additional band structures).

We start with the domains of the helical trilayer. Within each local domain of the relaxed supermoir\'e structure illustrated in Fig.~\ref{fig:hel_stru}, the $1$--$2$ and $2$--$3$ moir\'e lattices locally align their orientations. We therefore model each domain by slightly expanding the lattice constant of the middle layer so that the $K_1$, $K_2$, and $K_3$ valleys of the three layers are aligned in momentum space, as illustrated in Fig.~\ref{fig:hel_band}(a). We further fix the twist angle pair at $\theta_{12}=\theta_{23}=3.67^\circ$, for which the system becomes commensurate and thus suitable for \emph{ab initio} calculation. The $1$--$2$ and $2$--$3$ moir\'e patterns can still differ by an in-plane shift, allowing us to access different local domains by varying this shift. We then perform lattice relaxation of a given shift. As an example, the resulting moir\'e lattice for the $\alpha\beta$ domain in htWSe$_2$, characterized by the interlayer distance $d_z^{(23)}$ between layers 2 and 3, is shown in Fig.~\ref{fig:hel_band}(b).

Figures~\ref{fig:hel_band}(c) and \ref{fig:hel_band}(d) show the moir\'e valence band structures for htWSe$_2$ in the $\alpha\beta$ and $\beta\alpha$ domains, respectively. The black solid and dashed curves denote states from the $K$ and $K'$ valleys, which have opposite spins due to spin-valley locking and are related by time-reversal symmetry, while the red curves originate from the $\Gamma$ valley. 
The two topmost valence bands are energetically isolated and form a time-reversal pair derived from the $K$ and $K'$ valleys, with parabolic dispersions around $\kappa$, $\gamma$, and $\kappa'$ inherited from the valence-band maxima of the three constituent layers.
The global band maximum occurs at $\gamma$ for the $\alpha\beta$ domain and at $\kappa/\kappa'$ for the $\beta\alpha$ domain. The $K$-valley-derived topmost valence band has Chern number $C_K=+1$ in the $\alpha\beta$ domain and $C_K=-1$ in the $\beta\alpha$ domain, while the time-reversal-related $K'$-valley bands carry opposite Chern numbers, $C_{K'}=-C_K$. The spatial domain reconstruction in Fig.~\ref{fig:hel_stru}(a), together with the domain-dependent Chern numbers, gives rise to a supermoir\'e topological mosaic~\cite{guerci2024chern}, with larger $C_K=+1$ $\alpha\beta$ domains and smaller $C_K=-1$ $\beta\alpha$ domains.

Figures~\ref{fig:hel_band}(e) and \ref{fig:hel_band}(f) show the moir\'e band structures of htMoTe$_2$ in the $\alpha\beta$ and $\beta\alpha$ domains, respectively, which share similarities with but also exhibit distinct features from those of htWSe$_2$. Similar to htWSe$_2$, the isolated topmost valence bands originate from the $K$ and $K'$ valleys, with the global band maximum located at $\gamma$ for the $\alpha\beta$ stacking and at $\kappa/\kappa'$ for the $\beta\alpha$ stacking. However, htMoTe$_2$ exhibits two notable differences: (i) in the $\beta\alpha$ domain, the $K$-valley-derived topmost band retains $C_K=-1$ but becomes substantially flatter, with a bandwidth of $3.7$~meV compared with $13.8$~meV in htWSe$_2$, due to the larger effective mass of MoTe$_2$; (ii) in the $\alpha\beta$ domain, the topmost bands are topologically trivial with $C_K=0$. These results reveal that htMoTe$_2$ forms a supermoir\'e topological mosaic consisting of large hexagonal-like $C_K=-1$ $\beta\alpha$ domains and small three-pointed-star-shaped $C_K=0$ $\alpha\beta$ domains. Unlike htWSe$_2$, where only the topmost $K$/$K'$-valley-derived valence bands are isolated from the $\Gamma$-valley-derived bands, htMoTe$_2$ features a broader set of isolated $K$/$K'$-valley-derived valence bands. Several of these top valence bands exhibit nonzero Chern numbers, including higher-Chern bands, as indicated in Fig.~\ref{fig:hel_band}.

We next consider the alternating trilayer with $\theta_{12}=-\theta_{23}$, for which the $1$--$2$ and $2$--$3$ moir\'e lattices share an identical period [Fig.~\ref{fig:alt_band}(a)]. We choose the commensurate twist angle $\theta_{12}=-\theta_{23}=3.89^\circ$. Relative in-plane shifts between the two moir\'e lattices generate distinct local domains, including $\alpha\alpha^\prime$, $\alpha\beta^\prime$, and $\beta\alpha^\prime$. 
The $\alpha\alpha^\prime$ domain is invariant under the mirror reflection $M_z$ about layer 2, while the $\alpha\beta^\prime$ and $\beta\alpha^\prime$ domains are interchanged by $M_z$. The relaxed $\alpha\alpha^\prime$ domain of atWSe$_2$, characterized by the interlayer distance $d_z^{(23)}$, is shown in Fig.~\ref{fig:alt_band}(b).

Figures~\ref{fig:alt_band}(c) and \ref{fig:alt_band}(d) show the moir\'e valence band structures for atWSe$_2$ in the $\alpha\alpha^\prime$ and $\alpha\beta^\prime$ domains, respectively. The topmost $K$-valley-derived and $\Gamma$-valley-derived bands overlap in energy near the $\gamma$ point. For the $\alpha\alpha^\prime$ domain, the Hamiltonian can be block diagonalized into mirror-even and mirror-odd sectors using the $M_z$ symmetry~\cite{li2019electronics,khalaf2019magic}. The $K$-valley topmost band in the $\alpha\alpha^\prime$ domain originates from the mirror-even sector, carrying a Chern number of $C_K=-1$. In contrast, the $K$-valley topmost band in the $\alpha\beta^\prime$ domain is topologically trivial with $C_K=0$, and the $\beta\alpha^\prime$ domain shares the same trivial topology due to its mirror relationship with $\alpha\beta^\prime$. The results here agree with a recent \emph{ab initio} study~\cite{fan2026alternating}, while additionally accounting for the $\Gamma$ valley alongside $K$ and $K^\prime$.

The moir\'e valence band structures of atMoTe$_2$, shown in Figs.~\ref{fig:alt_band}(e) and (f), are markedly different from those of atWSe$_2$. Several $K$/$K'$-valley-derived valence bands are energetically isolated from the $\Gamma$-valley-derived bands, forming narrow moir\'e bands. In the $\alpha\alpha^\prime$ domain, the top three $K$-valley-derived bands, ordered by decreasing energy, belong to the mirror-even, mirror-odd, and mirror-even sectors, respectively, and carry Chern numbers $-1$, $0$, and $1$. In the $\alpha\beta^\prime$ domain, the top two $K$-valley-derived bands are split by a tiny gap and together carry a total Chern number of $-1$, while the third band is energetically isolated and carries Chern number $1$.

\textit{Discussion.—}
Very recent experiments have begun to reveal a rich landscape of correlated and topological phenomena in twisted trilayer TMDs. In htWSe$_2$ with $\theta_{12}=\theta_{23}=3.9^\circ$, ferromagnetic states accompanied by a non-quantized anomalous Hall response appear at filling factor $\nu=-1$ (one hole per moir\'e cell)~\cite{wang2026supermoire}. In atWSe$_2$ with $\theta_{12}=-\theta_{23}=3.87^\circ$, correlated insulating states and superconductivity emerge near $\nu=-1/2$ and $-1$~\cite{jeong2026superconductivity}. In atMoTe$_2$ with $\theta_{12}=-\theta_{23}=3.9^\circ$, signatures of electrically tunable integer and fractional Chern insulators emerge~\cite{beach2026electrically}.

Our calculations suggest that the relaxation-induced supermoir\'e domain structure plays an important role in shaping these observed phenomena. 
In helical trilayers, the dominant $\alpha\beta$ and $\beta\alpha$ domains host topological bands with distinct valley Chern numbers, forming a topological mosaic. The resulting domain walls between regions of different topology support topological boundary states, which, in combination with spontaneous valley polarization, contribute to the experimentally observed non-quantized anomalous Hall effect. A quantitative description requires treating transport through the domain-wall network~\cite{bhattacharjee2026mesoscopic}.
In alternating trilayers, an inevitable small difference between $\theta_{12}$ and $-\theta_{23}$ gives rise to a supermoir\'e pattern, while lattice relaxation drives the system to be dominated by $\alpha\alpha^\prime$ domains, as illustrated in Fig.~\ref{fig:alt_stru}. The physics of the $\alpha\alpha^\prime$ domains can therefore play a dominant role in determining the electronic properties. Our results provide a microscopic starting point for investigating the many-body physics observed experimentally.

Interference between moir\'e patterns of distinct length scales was recently visualized in twisted trilayer WS$_2$~\cite{xiao2025deep}, motivating future imaging of the relaxed supermoir\'e structures with comparable $|\theta_{12}|$ and $|\theta_{23}|$. 
Such measurements should be feasible in supermoir\'e TMDs, given that imaging techniques of twisted trilayer graphene are rapidly advancing~\cite{hoke2026imaging,craig2024local,li2025engineering,hao2024robust,wen2023moire}. Unlike helical trilayer graphene, helical trilayer TMDs lack $C_{2z}$ symmetry, making the $\alpha\beta$ and $\beta\alpha$ domains inequivalent and yielding an asymmetric reconstruction, which calls for experimental examination. The domain-dependent valley Chern numbers could be determined by tracking St\v{r}eda trajectories in scanning tunneling spectroscopy maps of carrier density versus out-of-plane magnetic field.

\textit{Acknowledgments.—}
We thank Quansheng Wu and Yan Zhang for valuable discussions.
This work was supported by the National Key Research and Development Program of China (Grants No.~2021YFA1401300 and No.~2022YFA1402400), and the National Natural Science Foundation of China (Grants No.~12274333 and No.~12550404). The numerical calculations in this paper have been performed on the supercomputing system in the Supercomputing Center of Wuhan University.

\bibliography{ref}

\end{document}

\title{Supplemental Material of ``Supermoir\'e Reconstruction and Topological Mosaics in Twisted Trilayer WSe$_2$ and MoTe$_2$"}

\author{Hai Meng}
%\email{menghai@whu.edu.cn}
\affiliation{School of Physics and Technology, Wuhan University, Wuhan 430072, China}
\author{Yang Xu}
%\email{yang.xu@iphy.ac.cn}
\affiliation{Beijing National Laboratory for Condensed Matter Physics, Institute of Physics, Chinese Academy of Sciences, Beijing 100190, China}
\affiliation{School of Physical Sciences, University of Chinese Academy of Sciences, Beijing 100049, China}
\author{Fengcheng Wu}
\email{wufcheng@whu.edu.cn}
\affiliation{School of Physics and Technology, Wuhan University, Wuhan 430072, China}
\affiliation{Wuhan Institute of Quantum Technology, Wuhan 430206, China}

\maketitle
This Supplemental Material includes the following four
sections: (1) construction of twisted trilayer moir\'e and supermoir\'e lattices with rigorous periods, (2) details of the machine learning force fields (MLFFs) training and \emph{ab initio} calculations, (3) projection of the Hamiltonian through the truncated atomic plane wave method (TAPW), and (4) supplemental results including relaxation, Berry curvature, wave functions, and band structures.

\section{Construction of periodic structures}
\label{sec:stru}
In order to relax and perform \emph{ab initio} calculations for twisted trilayer TMDs, the lattice structure should have rigorous periods. In the following, we present the methodology for constructing the twisted trilayer structures considered in the main text.

We define the two primitive lattice vectors of non-rotated monolayer TMD as $\bm{a}_1=a(1,0)$ and $\bm{a}_2=a(\frac{1}{2},\frac{\sqrt{3}}{2})$, where $a$ is the monolayer lattice constant.

Commensuration between the twisted layers $l$ and $l^\prime$ is achieved when they share the common lattice vector, defining a moir\'e lattice vector $\bm{a}^{(l^\prime l)}_M$ given by
\begin{equation}
\bm{a}^{(l^\prime l)}_\text{M} = m\bm{a}^{(l)}_1+n\bm{a}^{(l)}_2=m^\prime\bm{a}^{(l^\prime)}_1+n^\prime\bm{a}^{(l^\prime)}_2,
\end{equation}
where $m,n,m^\prime,n^\prime\in\mathbb{N}$ and $\bm{a}^{(l)}_i=R(\theta^{(l)})\lambda^{(l)}\bm{a}_i$. $R(\theta^{(l)})$ is the rotation matrix and $\lambda^{(l)}$ is the scaling factor. The commensurate twist angle between the two layers, the scaling factor ratio, and the moir\'e length are given by
\begin{align}
&\theta_c=\cos^{-1}{\frac{mm^\prime+nn^\prime+\frac{1}{2}(mn^\prime+m^\prime n)}{\sqrt{m^2+n^2+mn}\sqrt{m^{\prime2}+n^{\prime2}+m^\prime n^\prime}}},\notag\\
&\frac{\lambda^{(l^\prime)}}{\lambda^{(l)}}=\frac{\sqrt{m^2+n^2+mn}}{\sqrt{m^{\prime2}+n^{\prime2}+m^\prime n^\prime}}\notag\\
&|\bm{a}^{(l^\prime l)}_\text{M}|=\lambda^{(l)}\sqrt{m^2+n^2+mn}a.
\end{align}
We let $m^\prime=m-1$ and $n^\prime=n+1$ in the following.

For supermoir\'e helical trilayers with equal twist, i.e. $(\theta_{12},\theta_{23})= (\theta,\theta)$, the supermoir\'e structure has a rigorous period when $\theta$ is a commensurate angle for twisted bilayer. We choose $(m,n)=(9,8)$ for $(l,l^\prime)=(2,1)$ and $(l,l^\prime)=(3,2)$ so that $\theta\approx3.89^\circ$ and $\lambda^{(1)}=\lambda^{(2)}=\lambda^{(3)}=1$. The two moir\'e lattice vectors $\bm{a}^{(12)}_\text{M}$ and $\bm{a}^{(23)}_\text{M}$ have equal periods and a relative twist angle of $\theta\approx3.89^\circ$. The supermoir\'e period is $a_\text{SM}=217a$.

For moir\'e structures in the commensurate domains $\alpha\alpha$, $\alpha\beta$, and $\beta\alpha$ of helical trilayer, we choose $(m,n)=(9,9)$ for $(l,l^\prime)=(2,1)$ and $(m,n)=(10,8)$ for $(l,l^\prime)=(3,2)$ so that $\theta_{12}=\theta_{23}\approx3.67^\circ$ and $\lambda^{(1)}=\lambda^{(3)}=1$, $\lambda^{(2)}\approx1.0021$. The two moir\'e lattice vectors coincide as $\bm{a}^{(12)}_\text{M}=\bm{a}^{(23)}_\text{M}$. The moir\'e period is $a_\text{M}=2\sqrt{61}a$. We can also construct commensurate helical trilayers for larger twist angles. For $\theta_{12}=\theta_{23}\approx4.13^\circ$, we have $(m,n)=(8,8)$ for $(l,l^\prime)=(2,1)$ and $(m,n)=(9,7)$ for $(l,l^\prime)=(3,2)$ so that $\lambda^{(1)}=\lambda^{(3)}=1$, $\lambda^{(2)}\approx1.0026$ and $a_\text{M}=\sqrt{193}a$. For $\theta_{12}=\theta_{23}\approx4.72^\circ$, $(m,n)=(7,7)$ for $(l,l^\prime)=(2,1)$ and $(m,n)=(8,6)$ for $(l,l^\prime)=(3,2)$ so that $\lambda^{(1)}=\lambda^{(3)}=1$, $\lambda^{(2)}\approx1.0034$ and $a_\text{M}=2\sqrt{37}a$.

The supermoir\'e helical trilayers with equal twist $\theta_{12}=\theta_{23}= \theta$ have $D_3$ point group symmetry, which is generated by threefold rotation $C_{3z}$ around the $z$ axis and twofold rotation $C_{2y}$ around the $y$ axis. The $C_{2y}$ symmetry exchanges layer 1 and 3. The domains $\alpha\alpha$, $\alpha\beta$, and $\beta\alpha$ also respect the $D_3$ point group symmetry. 

For supermoir\'e alternating trilayers, we choose $(m,n)=(15,15)$ for $(l,l^\prime)=(2,1)$ and $(m,n)=(18,18)$ for $(l,l^\prime)=(2,3)$ so that $\theta_{12}\approx2.20^\circ$, $\theta_{23}\approx-1.84^\circ$ and $\lambda^{(1)}\approx 0.9998$, $\lambda^{(2)}\approx1.0005$, $\lambda^{(3)}=1$. The two moir\'e lattices are orientationally aligned and have $6\bm{a}^{(12)}_\text{M}=5\bm{a}^{(23)}_\text{M}$. The supermoir\'e period is $a_\text{SM}=5\sqrt{973}a$.

For moir\'e structures in the commensurate domains $\alpha\alpha^\prime$, $\alpha\beta^\prime$, and $\beta\alpha^\prime$ of alternating trilayer, we choose $(m,n)=(9,8)$ for $(l,l^\prime)=(2,1)$ and $(l,l^\prime)=(2,3)$ so that $\theta_{12}=-\theta_{23}\approx3.89^\circ$ and $\lambda^{(1)}=\lambda^{(2)}=\lambda^{(3)}=1$. The two moir\'e lattice vectors coincide as $\bm{a}^{(12)}_\text{M}=\bm{a}^{(23)}_\text{M}$. The moir\'e period is $a_\text{M}=\sqrt{217}a$.

The $\alpha\alpha^\prime$ structure with $\theta_{12}=-\theta_{23}$ exhibits $C_{3h}$ point group symmetry, which is generated by threefold rotation $C_{3z}$ around $z$ axis and mirror reflection $M_z$ with respect to the middle layer. The $\alpha\beta^\prime$ and $\beta\alpha^\prime$ structures with $\theta_{12}=-\theta_{23}$ are related by the $M_z$ operation.

\begin{figure}[t]
    \centering
    \includegraphics[width=1.\columnwidth]{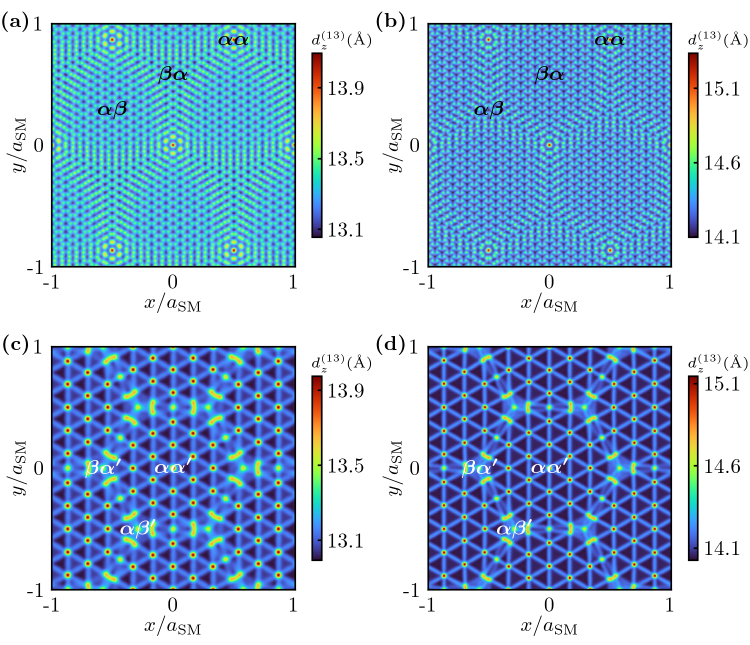}
    \caption{(a,b) Interlayer distance $d^{(13)}_z$ of relaxed htWSe$_2$ and htMoTe$_2$ with $\theta_{12}=\theta_{23}=3.89^\circ$, respectively. (c,d) Interlayer distance $d^{(13)}_z$ of relaxed atWSe$_2$ and atMoTe$_2$ with $(\theta_{12},\theta_{23})=(2.20^\circ,-1.84^\circ)$, respectively. $a_{\text{SM}}$ denotes the supermoir\'e period.} 
    \label{fig:dz}
\end{figure}

\section{Calculation details}
We use the deep learning package DeePMD-kit~\cite{wang2018dpmd1,zeng2023dpmd2} to construct MLFFs by fitting neural networks to the \emph{ab initio} data.
The cutoff radius for atomic neighbor searching is 11.0 \text{\AA} for trilayer WSe$_2$ and 13.0 \text{\AA} for trilayer MoTe$_2$. The difference in the choice of the cutoff radius is based on the larger interlayer distance in trilayer MoTe$_2$. The MLFFs are trained on the generated dataset by one million steps for WSe$_2$ and two million steps for MoTe$_2$ to minimize the loss function which includes energy, force, and virial contributions. For WSe$_2$, the final energy, virial, and force root mean squared error (RMSE) of the training set converge to $0.3$ meV/atom, $4.0$ meV, and $32.6$ meV/\text{\AA}, respectively. For MoTe$_2$, the final energy, virial, and force RMSE of the training set converge to $0.1$ meV/atom, $2.8$ meV, and $22.5$ meV/\text{\AA}, respectively. The convergence criterion is similar to that of the twisted bilayer TMD case~\cite{zhang2024polarization}.

\begin{figure}[t]
    \centering
    \includegraphics[width=1.\columnwidth]{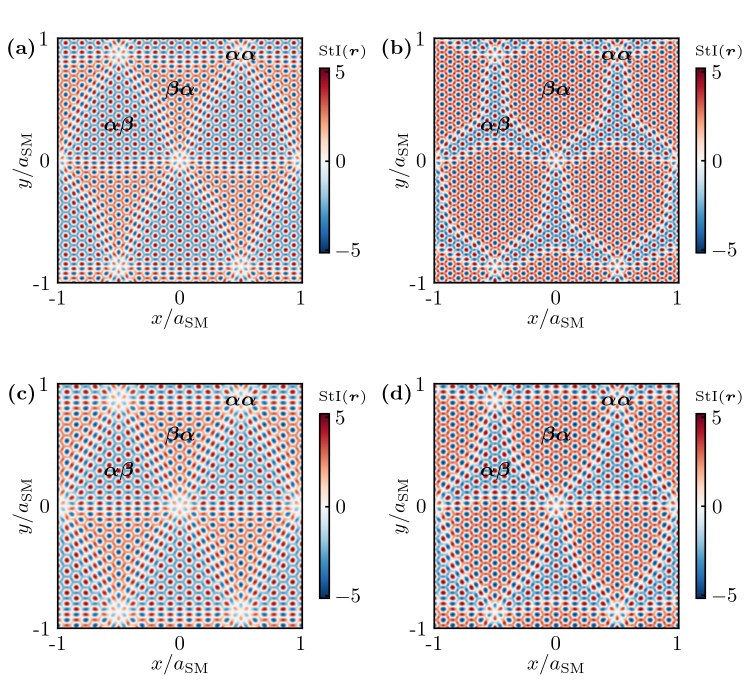}
    \caption{(a,b) Spatial pattern of relaxed htWSe$_2$ and htMoTe$_2$ with $\theta_{12}=\theta_{23}=4.41^\circ$, respectively. (c,d) Spatial pattern of relaxed htWSe$_2$ and htMoTe$_2$ with $\theta_{12}=\theta_{23}=5.09^\circ$, respectively.} 
    \label{fig:sti}
\end{figure}

\emph{Ab initio} calculations are performed with the DFT-based software ABACUS~\cite{chen2010abacus1,li2016abacus2}. We use the Perdew–Burke–Ernzerhof (PBE) exchange-correlation functional, norm-conserving pseudopotentials, and corresponding optimized double-zeta plus polarization (DZP) numerical atomic orbital (NAO) basis sets in all calculations. 
Van der Waals interactions are accounted for using Grimme's DFT-D3 dispersion correction with zero damping~\cite{grimme2010d3} for the WSe$_2$ case, and Grimme's DFT-D2 dispersion correction~\cite{grimme2006d2} for the MoTe$_2$ case.
For MLFF dataset generation, spin-orbit coupling (SOC) is not included, as it has a negligible effect on atomic forces. 
The WSe$_2$ MLFF dataset is generated using SG15 optimized norm-conserving Vanderbilt (ONCV) multi-projector pseudopotentials~\cite{schlipf2015optimization}, while the MoTe$_2$ MLFF dataset is generated using Dojo norm-conserving fully relativistic pseudopotentials~\cite{vansetten2018pseudo}.

For electronic Hamiltonian calculations of relaxed moir\'e structures, we use Dojo norm-conserving fully relativistic pseudopotentials for both WSe$_2$ and MoTe$_2$. We first perform a non-SOC self-consistent calculation to obtain the charge density, and then include SOC effects without further iterating the charge density, obtaining the Hamiltonian and overlap matrices in the non-orthogonal NAO basis.

\section{TAPW Method}
The DZP NAO basis sets we choose for the electronic calculations are specified as M-$4s2p2d1f$ and X-$2s2p2d1f$ where M (X) indicates the metal W/Mo (chalcogen Se/Te) atoms. The notation $4s2p2d1f$ for M atoms indicates $4$ $s$-orbitals, $2$ sets of $p$-orbitals, $2$ sets of $d$-orbitals, and $1$ set of $f$-orbitals, yielding a total of $27$ atomic orbitals per M atom. The notation $2s2p2d1f$ for X atoms indicates $2$ $s$-orbitals, $2$ sets of $p$-orbitals, $2$ sets of $d$-orbitals, and $1$ set of $f$-orbitals, yielding a total of $25$ atomic orbitals per X atom. 

\begin{figure*}[t]
    \centering
    \includegraphics[width=1.\textwidth]{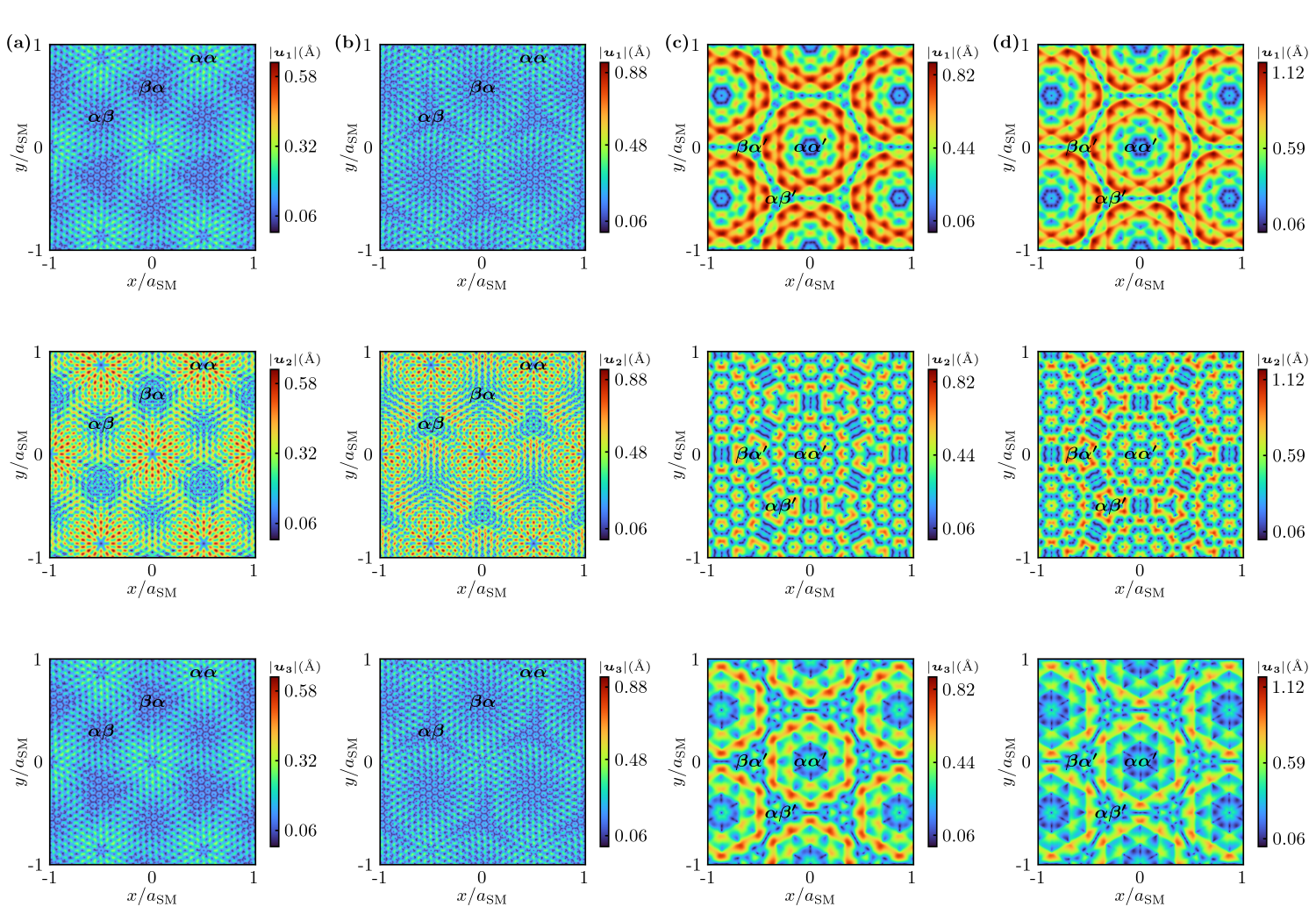}
    \caption{(a,b) In-plane displacement $|\bm{u}_l|$ of relaxed htWSe$_2$ and htMoTe$_2$ with $\theta_{12}=\theta_{23}=3.89^\circ$, respectively. (c,d) In-plane displacement $|\bm{u}_l|$ of relaxed atWSe$_2$ and atMoTe$_2$ with $(\theta_{12},\theta_{23})=(2.20^\circ,-1.84^\circ)$, respectively. The first, second, and third rows correspond to the displacement of layer $1$, $2$, and $3$, respectively.} 
    \label{fig:u}
\end{figure*}

As an example, in the $\alpha\alpha$ commensurate domain (or equivalently, $\alpha\beta$ or $\beta\alpha$) of the helical trilayer TMD with $\theta_{12}=\theta_{23}=3.67^\circ$, the \emph{ab initio} Hamiltonian matrix (including SOC) has a dimension of $112574$. Solving the high-dimensional eigenvalue problem for such systems can be computationally challenging. Here, we adopt the TAPW method~\cite{miao2023tapw} by projecting the \emph{ab initio} Hamiltonians and overlap matrices onto $K$, $K^\prime$, and $\Gamma$ valleys to reduce the dimensions of the eigenvalue problem.

The NAO Bloch basis is given by
\begin{equation}
\label{eq:bloch}
\psi_{li\alpha}(\bm{k})=\frac{1}{\sqrt{N_S}}\sum_{\bm{R}}e^{i\bm{k}\cdot(\bm{R}+\bm{\tau}_{li\alpha})}\phi_\alpha(\bm{r}-\bm{R}-\bm{\tau}_{li\alpha}),
\end{equation}
where $N_S$ is the number of moir\'e supercells, $\bm{R}$ is the lattice vector of the supercell, $li\alpha$ denotes the $\alpha$-th atomic orbital of the $i$-th monolayer primitive cell of layer $l$. $\alpha\equiv snlm\sigma$, where $s=\text{M},\text{X1},\text{X2}$ denotes the atoms in a primitive cell, $nlm$ denote the orbital quantum numbers and $\sigma=\uparrow,\downarrow$ is the spin index. $\bm{\tau}_{li\alpha}$ is the displacement of orbital $li\alpha$ in a moir\'e supercell and $\phi_\alpha(\bm{r}-\bm{R}-\bm{\tau}_{li\alpha})$ is the atomic orbital wavefunction. 

We now define the \textit{atomic plane wave} basis as
\begin{equation}
\tilde{\psi}_{lm\alpha}(\bm{k})=\frac{1}{\sqrt{N_SN_l}}\sum_{\bm{R},i}e^{i(\bm{k}+\bm{G}_{lm})\cdot(\bm{R}+\bm{\tau}_{li\alpha})}\phi_\alpha(\bm{r}-\bm{R}-\bm{\tau}_{li\alpha}),
\end{equation}
where $N_l$ is the number of monolayer primitive cell of layer $l$ in a moir\'e supercell and $\bm{G}_{lm}$ is the $m$-th moir\'e reciprocal lattice vector of layer $l$. The atomic plane wave basis and the NAO Bloch basis are related by
\begin{equation}
\label{eq:relation}
\tilde{\psi}_{lm\alpha}(\bm{k}) = \frac{1}{\sqrt{N_l}}\sum_ie^{i\bm{G}_{lm}\cdot\bm{\tau}_{li\alpha}}\psi_{li\alpha}
\end{equation}

Using the relation in Eq.~\eqref{eq:relation}, we can obtain the Hamiltonian and overlap matrix in the atomic plane wave basis as
\begin{align}
\label{eq:TAPW}
&\tilde{H}_{l m \alpha,l^\prime n \beta}(\bm{k}) = \sum_{ij}X^*_{li\alpha,lm\alpha} H_{li\alpha,l^\prime j \beta}(\bm{k})X_{l^\prime j \beta,l^\prime n \beta},\notag\\
&\tilde{S}_{l m \alpha,l^\prime n \beta}(\bm{k}) = \sum_{ij}X^*_{li\alpha,lm\alpha} S_{li\alpha,l^\prime j \beta}(\bm{k})X_{l^\prime j \beta,l^\prime n \beta},\notag\\
&X_{li\alpha,lm\alpha} = \frac{e^{i\bm{G}_{lm}\cdot\bm{\tau}_{li\alpha}}}{\sqrt{N_l}},
\end{align}
where $H(\bm{k})$ and $S(\bm{k})$ are the \emph{ab initio} Hamiltonian and overlap matrix in NAO Bloch basis.

\begin{figure}[t]
    \centering
    \includegraphics[width=1.\columnwidth]{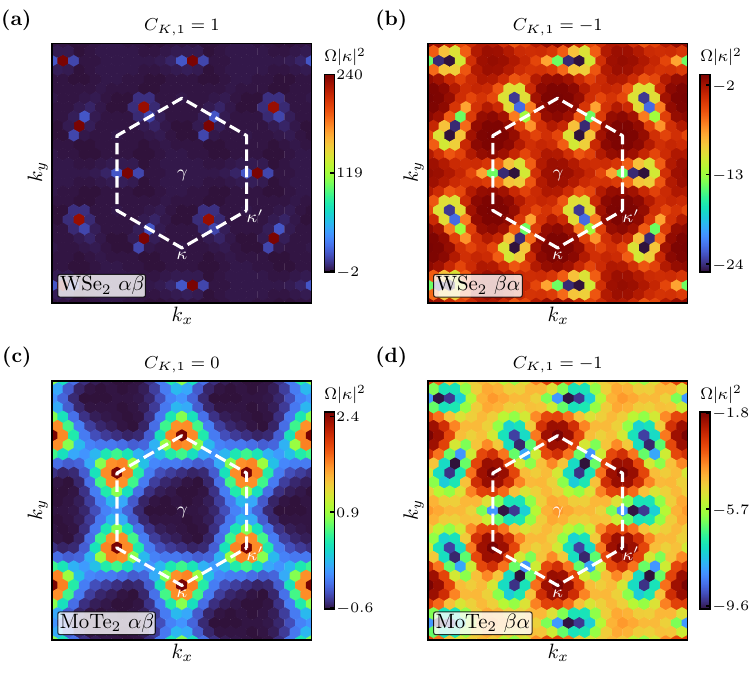}
    \caption{Berry curvature of the topmost $K$ valley moir\'e bands of helical trilayer TMDs. White dashed hexagon denotes the moir\'e Brillouin zone. Panels (a-d) correspond to htWSe$_2$ in the $\alpha\beta$ (a) and $\beta\alpha$ (b) domains, and htMoTe$_2$ in the $\alpha\beta$ (c) and $\beta\alpha$ (d) domains. In (a)-(d), $\theta_{12}=\theta_{23}=3.67^\circ$.} 
    \label{fig:hel_bc}
\end{figure}

Considering that the low energy moir\'e bands mainly originate from the $K$, $K^\prime$, and $\Gamma$ valleys, we can truncate the number of atomic plane waves by selecting $\bm{G}_{lm}$ within a certain distance around the valley $\zeta=K,K^\prime$ or $\Gamma$ to obtain the $\zeta$-valley-projected $\tilde{H}^{(\zeta)}(\bm{k})$ and $\tilde{S}^{(\zeta)}(\bm{k})$. In practice, we choose $37$ atomic plane waves for $\tilde{H}^{(\zeta)}(\bm{k})$ and $\tilde{S}^{(\zeta)}(\bm{k})$. Through the TAPW method, the valley-projected Hamiltonian for the helical trilayer TMD with $\theta_{12}=\theta_{23}=3.67^\circ$ in the $\alpha\alpha$ commensurate domain (equivalently, $\alpha\beta$ or $\beta\alpha$) is reduced from $112574$ to $17094$.

For the calculation of Berry curvature and Chern numbers, we can orthogonalize the Hamiltonian by,
\begin{equation}
\tilde{\mathcal{H}}^{(\zeta)}(\bm{k}) = [\tilde{S}^{(\zeta)}(\bm{k})]^{-1/2}\tilde{H}^{(\zeta)}(\bm{k})[\tilde{S}^{(\zeta)}(\bm{k})]^{-1/2},
\end{equation}
which reduces the generalized eigenvalue problem to a standard eigenvalue problem. Berry curvatures and Chern numbers are calculated by the Fukui-Hatsugai-Suzuki method~\cite{fukui2005chern}. For a uniform momentum space $N\times N$ grid, the Berry flux through each plaquette is given by
\begin{align}
&F_{12}(\bm{k}_{ij}) = \arg\left[U_{1}(\bm{k}_{ij})U_{2}(\bm{k}_{ij}+\bm{\delta}_1)U_{1}^*(\bm{k}_{ij}+\bm{\delta}_2)U_{2}^*(\bm{k}_{ij})\right],\notag\\
&U_{\mu}(\bm{k}) = \langle u(\bm{k}) \vert u(\bm{k}+\bm{\delta}_\mu)\rangle, \quad \bm{\delta}_\mu=\bm{b}_\mu/N, \quad \mu=1,2 
\end{align}
where $\bm{b}_1$ and $\bm{b}_2$ are reciprocal lattice vectors, $\bm{k}_{ij}=i\bm{\delta}_1+j\bm{\delta}_2$ is the grid point, and $\lvert u(\bm{k})\rangle$ is the periodic part of the Bloch eigenstate. We use a counterclockwise momentum loop in the calculation of $F_{12}(\bm{k}_{ij})$. 

The Berry curvature is approximated as
\begin{equation}
\Omega(\bm{k}_{ij}) = \frac{F_{12}(\bm{k}_{ij})}{\Delta S_{\bm{k}}},
\end{equation}
where $\Delta S_{\bm{k}}$ is the area of one plaquette. Chern number is obtained by integrating the Berry curvature over the Brillouin zone,
\begin{equation}
C = \frac{1}{2\pi}\sum_{i,j}F_{12}(\bm{k}_{ij}).
\end{equation}

\begin{figure}[t]
    \centering
    \includegraphics[width=1.\columnwidth]{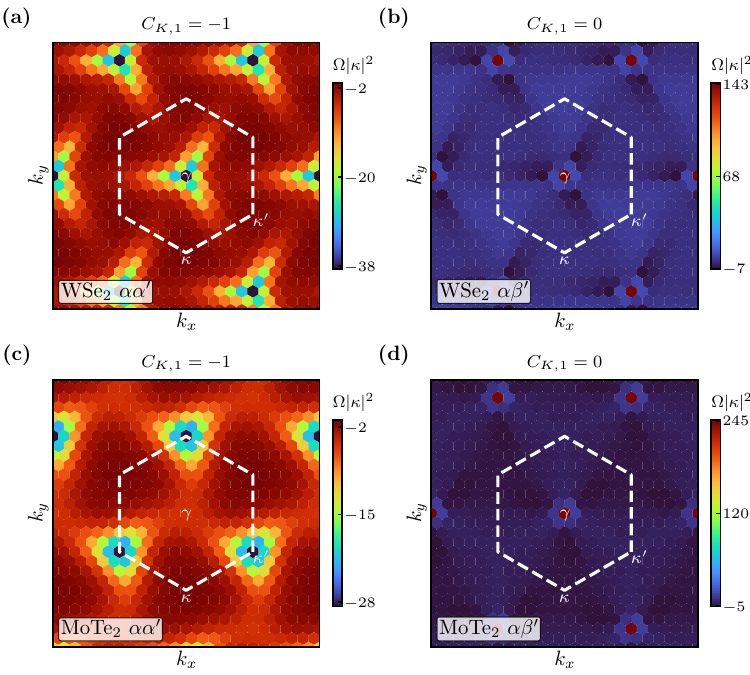}
    \caption{Berry curvature of the topmost $K$ valley moir\'e bands of alternating trilayer TMDs. White dashed hexagon denotes the moir\'e Brillouin zone. Panels (a-d) correspond to atWSe$_2$ in the $\alpha\alpha^\prime$ (a) and $\alpha\beta^\prime$ (b) domains, and atMoTe$_2$ in the $\alpha\alpha^\prime$ (c) and $\alpha\beta^\prime$ (d) domains. In (a)-(d), $\theta_{12}=-\theta_{23}=3.89^\circ$.} 
    \label{fig:alt_bc}
\end{figure}

\section{Supplemental results}
In Fig.~\ref{fig:dz}, we provide a complementary characterization of the spatial pattern of relaxed helical and alternating TMDs through the interlayer distance $d_z^{(13)}$. 
Figures~\ref{fig:dz}(a)--(b) and (c)--(d) show the same systems as Figs.~1(a)--(b) and 2(a)--(b) in the main text, respectively.

\begin{figure}[!htbp]
    \centering
    \includegraphics[width=1.\columnwidth]{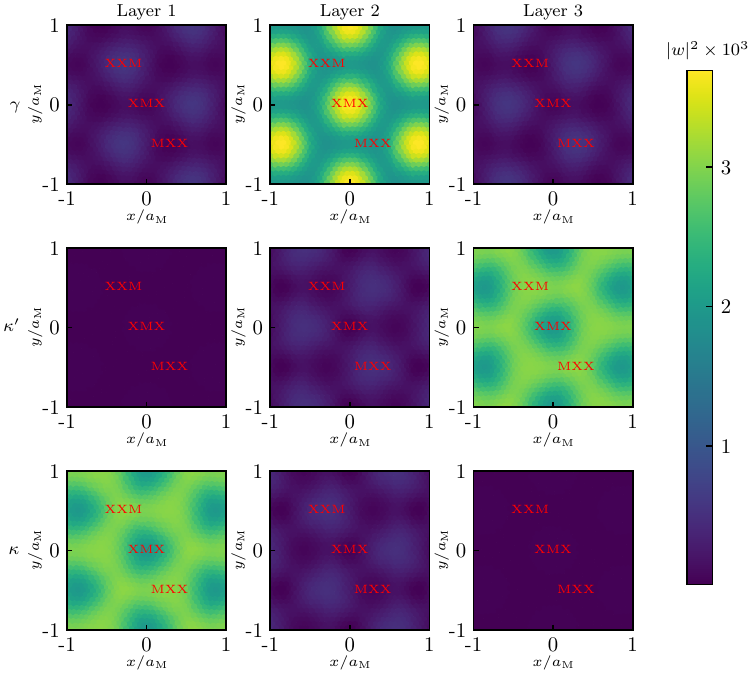}
    \caption{Real-space wavefunction of the $K$ valley topmost valence band eigenstate projected to each layer for $\alpha\beta$ htWSe$_2$. The first, second, and third rows correspond to the eigenstates at $\gamma$, $\kappa^\prime$, and $\kappa$ points respectively. The first, second, and third columns correspond to the components of layer $1$, $2$, and $3$, respectively.} 
    \label{fig:wf1}
\end{figure}

\begin{figure}[!htbp]
    \centering
    \includegraphics[width=1.\columnwidth]{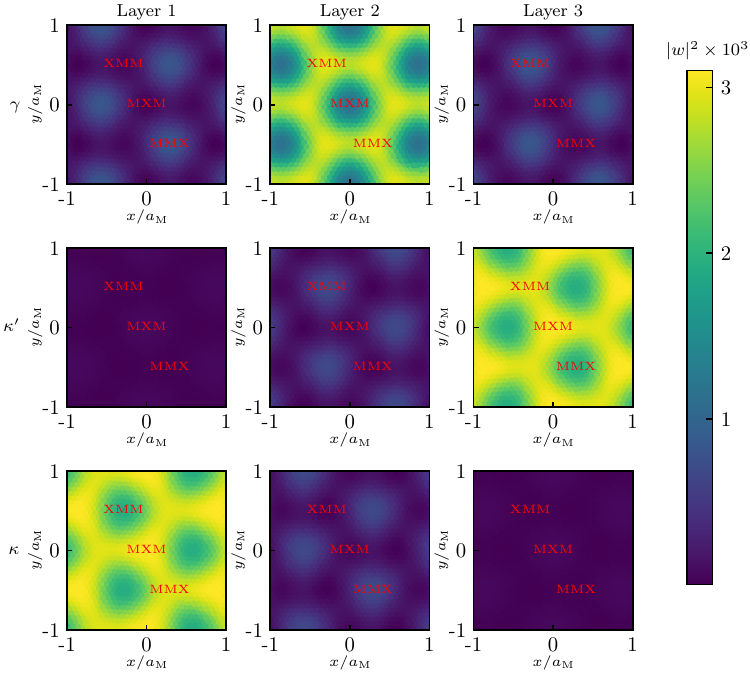}
    \caption{Real-space wavefunction of the $K$ valley topmost valence band eigenstate projected to each layer for $\beta\alpha$ htWSe$_2$. The first, second, and third rows correspond to the eigenstates at $\gamma$, $\kappa^\prime$, and $\kappa$ points respectively. The first, second, and third columns correspond to the components of layer $1$, $2$, and $3$ respectively.} 
    \label{fig:wf2}
\end{figure}

\begin{figure}[!htbp]
    \centering
    \includegraphics[width=1.\columnwidth]{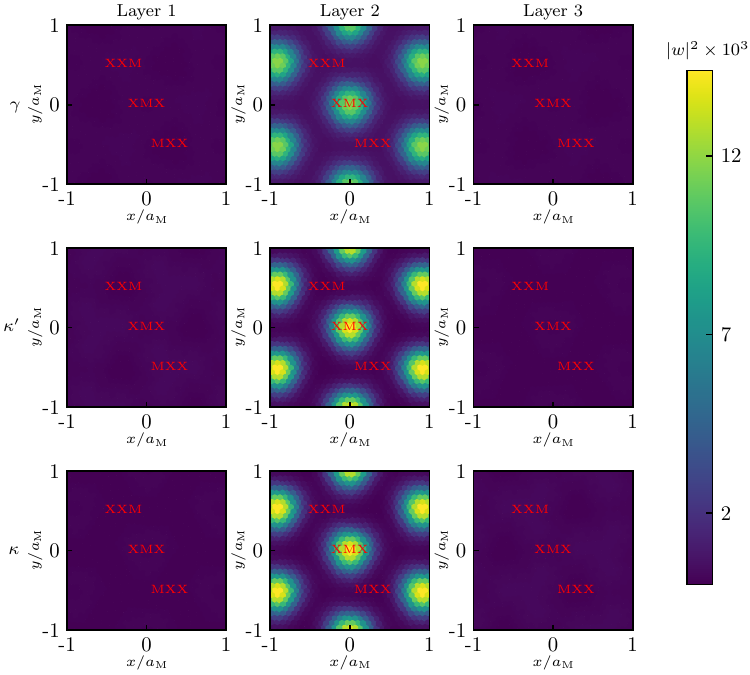}
    \caption{Real-space wavefunction of the $K$ valley topmost valence band eigenstate projected to each layer for $\alpha\beta$ htMoTe$_2$. The first, second, and third rows correspond to the eigenstates at $\gamma$, $\kappa^\prime$, and $\kappa$ points respectively. The first, second, and third columns correspond to the components of layer $1$, $2$, and $3$ respectively.} 
    \label{fig:wf3}
\end{figure}

\begin{figure}[!htbp]
    \centering
    \includegraphics[width=1.\columnwidth]{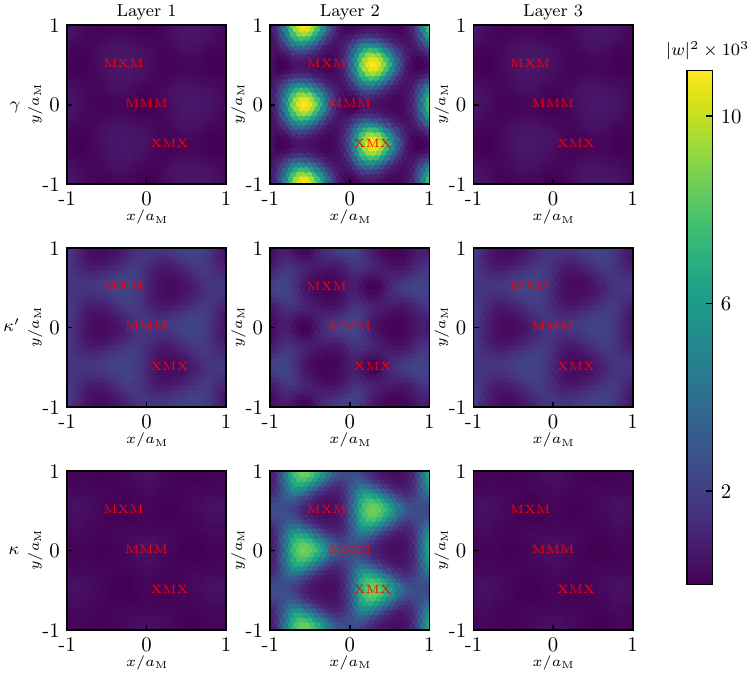}
    \caption{Real-space wavefunction of the $K$ valley topmost valence band eigenstate projected to each layer for $\alpha\alpha^\prime$ htMoTe$_2$. The first, second, and third rows correspond to the eigenstates at $\gamma$, $\kappa^\prime$, and $\kappa$ points respectively. The first, second, and third columns correspond to the components of layer $1$, $2$, and $3$ respectively.} 
    \label{fig:wf4}
\end{figure}

In Fig.~\ref{fig:sti}, we show the relaxed supermoir\'e structures of htWSe$_2$ and htMoTe$_2$ with twist angles of $4.41^\circ$ and $5.09^\circ$. The $\alpha\beta$ ($\beta\alpha$) domain is energetically favored in htWSe$_2$ (htMoTe$_2$), consistent with the domain preference in the $3.89^\circ$ helical trilayers discussed in the main text. In Tab.~\ref{tab:energy}, we list the energy difference $\Delta E$ between $\alpha\beta$ and $\beta\alpha$ commensurate structures at different twist angles (constructions are discussed in Sec.~\ref{sec:stru}). Here $\Delta E\equiv E_{\alpha\beta}-E_{\beta\alpha}$ where $E_{\alpha\beta}$ ($E_{\beta\alpha}$) is the energy per area for relaxed $\alpha\beta$ ($\beta\alpha$) commensurate moir\'e structures calculated using the MLFFs. The relative energy difference is consistent with the domain preference in relaxed supermoir\'e structures in Fig.~\ref{fig:sti} and Fig.~1 in the main text. Furthermore, we find that as the twist angle increases, the bending of the domain wall becomes weaker. This can be explained as follows. As shown in Tab.~\ref{tab:energy}, $\Delta E$ is insensitive to twist-angle variations. However, the supermoir\'e cell area decreases with increasing twist angle, which reduces the total energy difference between the $\alpha\beta$ and $\beta\alpha$ domains and leads to less bent domain walls.

\begin{table}[t]
    \centering
    \begin{tabular}{|c|c|c|c|}
        \hline
        & $3.67^\circ$ & $4.13^\circ$ & $4.72^\circ$ \\
        \hline
        htWSe$_2$ & -0.041 & -0.037 & -0.034 \\
        \hline
        htMoTe$_2$ & 0.073 & 0.077 & 0.070\\
        \hline
    \end{tabular}
    \caption{Energy difference per area $\Delta E$ (meV/\text{\AA}$^2$) between $\alpha\beta$ and $\beta\alpha$ commensurate structures. The first line lists the twist angle $\theta_{12}=\theta_{23}$. The construction of the commensurate structures is provided in Sec.~\ref{sec:stru}.}
    \label{tab:energy}
\end{table}

In Fig.~\ref{fig:u}, we present the multiscale relaxation of supermoir\'e structures through the in-plane relaxation displacement $|\bm{u}_1|$, $|\bm{u}_2|$, and $|\bm{u}_3|$ of the three layers. For helical trilayers in Figs.~\ref{fig:u} (a) and (b), $|\bm{u}_1|$ ($|\bm{u}_3|$) show clear features of moir\'e scale relaxation with small light-green hollow circles around the MM points of layer $2$ and $1$ ($3$), enlarging the XM/MX domains, while $|\bm{u}_2|$ shows characteristics of relaxation on the supermoir\'e scale with dark-red circle patterns around the $\alpha\alpha$ point, enlarging the $\alpha\beta$/$\beta\alpha$ domains. For alternating trilayers in Figs.~\ref{fig:u} (c) and (d), moir\'e scale relaxation can be observed in $|\bm{u}_2|$ with small light-red hollow circles around the MMM points, enlarging the XMX/MXM domains, while $|\bm{u}_1|$ and $|\bm{u}_3|$ show supermoir\'e relaxation features with dark-red ring-like patterns, enlarging the $\alpha\alpha^\prime$ domains.

Figures~\ref{fig:hel_bc} and~\ref{fig:alt_bc} show the Berry curvature distribution of the topmost valence band derived from the $K$-valley, corresponding to Figs.~3 and~4 in the main text, respectively.

\begin{figure}[b]
    \centering
    \includegraphics[width=1.\columnwidth]{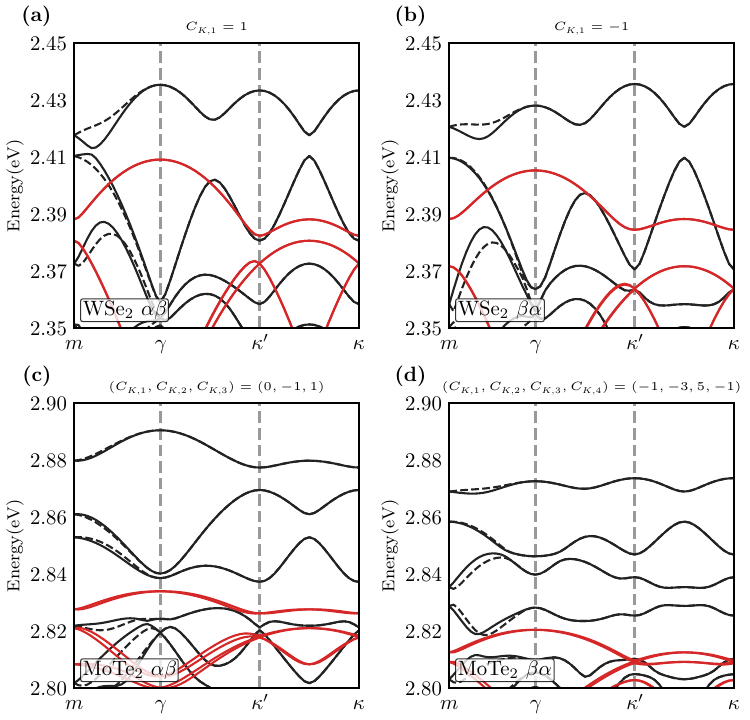}
    \caption{Moir\'e valence bands at the $K$ (black solid), $K^\prime$ (black dashed), and $\Gamma$ (red) valleys of helical trilayer TMDs.  Panels (a-d) correspond to htWSe$_2$ in the $\alpha\beta$ (a) and $\beta\alpha$ (b) domains, and htMoTe$_2$ in the $\alpha\beta$ (c) and $\beta\alpha$ (d) domains. In (a)-(d), $\theta_{12}=\theta_{23}=4.13^\circ$.} 
    \label{fig:bands1}
\end{figure}

\begin{figure}[b]
    \centering
    \includegraphics[width=1.\columnwidth]{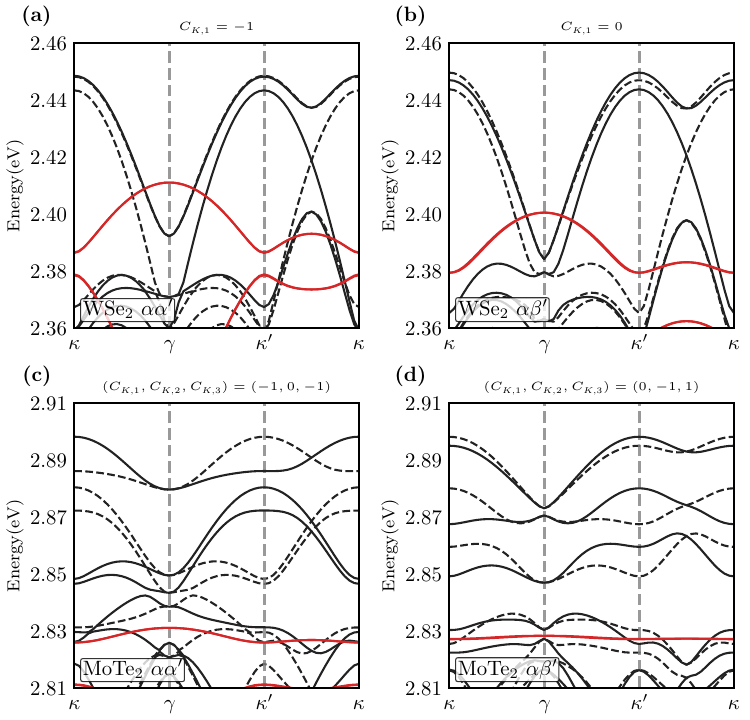}
    \caption{Moir\'e valence bands at the $K$ (black solid), $K^\prime$ (black dashed), and $\Gamma$ (red) valleys of alternating trilayer TMDs.  Panels (a-d) correspond to atWSe$_2$ in the $\alpha\alpha^\prime$ (a) and $\alpha\beta^\prime$ (b) domains, and atMoTe$_2$ in the $\alpha\alpha^\prime$ (c) and $\alpha\beta^\prime$ (d) domains. In (a)-(d), $\theta_{12}=-\theta_{23}=4.41^\circ$.} 
    \label{fig:bands2}
\end{figure}

In Figs.~\ref{fig:wf1}-\ref{fig:wf4}, we present the real-space wave function distributions of the $K$-valley topmost valence band states calculated directly from the \emph{ab initio} Hamiltonian for four moir\'e structures discussed in the main text.  The eigenstate of the generalized eigenvalue problem
\begin{equation}
H(\bm{k})\Psi(\bm{k})=E(\bm{k})S(\bm{k})\Psi(\bm{k}),
\end{equation}
is given by
\begin{equation}
\Psi(\bm{k})=\sum_{l,i,\alpha}c_{li\alpha}(\bm{k})\psi_{li\alpha}(\bm{k}),
\end{equation}
where $\psi_{li\alpha}(\bm{k})$ is the NAO Bloch basis defined in Eq.~\eqref{eq:bloch}, and $c_{li\alpha}(\bm{k})$ is the corresponding coefficient. The real-space distribution shown in the figures is obtained by plotting the weight of each metal atom,
\begin{equation}
w_{li}(\bm{k})=\sum_{\alpha\in\mathcal{O}_M}|c_{li\alpha}(\bm{k})|^2,
\end{equation}
where $\mathcal{O}_\text{M}=\{snlm\sigma|\, s=\text{M}\}$ is the set of orbital indices of the metal atom. We identify the $K$ ($K^\prime$)-valley state by applying the spin projection.

Figures~\ref{fig:wf1} and~\ref{fig:wf2} show the wavefunctions of $K$-valley topmost valence band in the $\alpha\beta$ and $\beta\alpha$ domains of htWSe$_2$, respectively. We see that the parabolic dispersions around $\kappa$, $\gamma$, and $\kappa'$ indeed originate from the $K$ valley valence band maxima of layers $1$, $2$, and $3$, respectively. The topmost valence band wavefunctions of $\beta\alpha$ htMoTe$_2$ share similar characteristics with those of $\beta\alpha$ htWSe$_2$. However, for the topologically trivial topmost valence band of $\alpha\beta$ htMoTe$_2$, as shown in Fig.~\ref{fig:wf3}, its wavefunctions are mainly contributed by the middle layer and reveal an effective electronic triangular lattice localized around the XMX point. 

For $\alpha\alpha^\prime$ atMoTe$_2$, the Hamiltonian can be block diagonalized into mirror-even and mirror-odd sectors using the $M_z$ reflection about layer 2. As shown in Fig.~\ref{fig:wf4}, we confirm that the topological topmost valence band originates from the mirror-even sector as the wavefunctions at $\kappa$, $\kappa^\prime$, and $\gamma$ contain layer 2 components. For $\kappa$ and $\gamma$ points, the wavefunction is concentrated at the XMX point in the middle layer. For $\kappa^\prime$ point, the wavefunction in the outer layers is concentrated at the MXM point. As in the twisted bilayer case, the topological physics originates from the hybridization of two orbitals forming a honeycomb lattice, analogous to Haldane-model physics \cite{wu2019topological}. In the bilayer case, these two orbitals are related by symmetry; in the trilayer case, however, one orbital is contributed by the middle layer and the other by the mirror-even combination of the two outer layers, so the two orbitals are no longer symmetry-equivalent and instead carry different onsite energies.

In Fig.~\ref{fig:bands1}, we show the moir\'e valence band structures for helical trilayer TMDs in the $\alpha\beta$ and $\beta\alpha$ domains at $\theta_{12}=\theta_{23}=4.13^\circ$. In Fig.~\ref{fig:bands2}, we show the moir\'e valence band structures for alternating trilayer TMDs in the $\alpha\alpha^\prime$ and $\alpha\beta^\prime$ domains at $\theta_{12}=-\theta_{23}=4.41^\circ$.  The top $K/K^\prime$-valley-derived valence bands of Figs.~\ref{fig:bands1} and~\ref{fig:bands2} exhibit similar dispersion shapes and identical Chern numbers compared to those of Figs.~3 and ~4 in the main text, because the twist angle differences are relatively small. As shown in Fig.~\ref{fig:bands2}(d), the top two $K(K^\prime)$-valley-derived valence bands of the $\alpha\beta^\prime$ domain in atMoTe$_2$ are separated by a larger band gap at $\theta_{12}=-\theta_{23}=4.41^\circ$ than at $\theta_{12}=-\theta_{23}=3.89^\circ$ [Fig.~4(f) in the main text].

\bibliography{ref}